\documentclass[12pt, onecolumn, article]{IEEEtran}

\usepackage[dvips]{graphicx}
\usepackage{amsmath}
\usepackage{amssymb}
\usepackage{amsfonts}

\usepackage{setspace}

\newcommand{\ignore}[1]{}
\newcommand{\HRule}{\rule{\linewidth}{0.5mm}}

\begin{document}

\title{Multiuser Switched Diversity Scheduling Schemes} 

\author{
\authorblockN{Mohammad Shaqfeh\,$^1$, Hussein Alnuweiri\,$^1$, and Mohamed-Slim Alouini\,$^2$}\\
\authorblockA{$^1$\, Texas A\&M University at Qatar, 
Doha, Qatar.\\ 
Email:  \{Mohammad.Shaqfeh, Hussein.Alnuweiri\}@qatar.tamu.edu \\[0.5ex]
$^2$\, King Abdullah University of Science and Technology (KAUST), \\
Thuwal, Mekkah Province, Saudi Arabia.\\ 
Email:  slim.alouini@kaust.edu.sa
\thanks{This paper was made possible by NPRP grant \# 08-577-2-241 from the Qatar
  National Research Fund (a member of Qatar Foundation). The statements made herein are solely the responsibility of the authors.}
\thanks{This work was partially presented in 7th IEEE Broadband Wireless Access Workshop (BWA2011) in Houston, Texas, US.} 
}}

\maketitle

\begin{abstract}
Multiuser switched-diversity scheduling schemes were recently proposed  
in order to overcome the heavy feedback requirements of conventional opportunistic scheduling 
schemes by applying a threshold-based, distributed, and ordered scheduling mechanism. 
The main idea behind these schemes is that slight reduction in 
the prospected multiuser diversity gains is an acceptable 
trade-off for great savings in terms of required channel-state-information feedback 
messages. In this work, we characterize the achievable rate region of
multiuser switched diversity systems and compare it with the rate region of
full feedback multiuser diversity systems. 
We propose also a novel proportional fair multiuser switched-based scheduling scheme 
and we demonstrate that it can be optimized using a practical 
and distributed method to obtain the feedback thresholds. 
We finally demonstrate by numerical examples that switched-diversity 
scheduling schemes operate within 0.3 bits/sec/Hz from the ultimate network capacity of 
full feedback systems in Rayleigh fading conditions.
\end{abstract}



\section{Introduction}
\label{sec:introduction}
The concept of \textit{multiuser diversity} (MUD) has been well studied in 
the literature, e.g. \cite[Chapter~6]{TsVi:05}, and exploited in the design of
channel-aware ``opportunistic'' scheduling schemes that control in a dynamic way how 
the users access the shared air-link resources in wireless systems. 
This concept was originally initiated in \cite{KnHu:95} where
it was shown that in order to maximize the sum capacity 
(bits/sec) of the network, we should always 
schedule the user with the best instantaneous channel quality.
The design of opportunistic schedulers 
has been further studied in the literature taking into consideration
key factors such as fairness among users and maintaining
the quality-of-service (QoS) constraints, e.g.~\cite{WaGiMa:2007}. 


In virtually all modern wireless communication systems, explicit training 
sequences (i.e.~pilot signals) are used to enable the receivers to measure
and/or predict (e.g. \cite{Du:2007})
the instantaneous channel conditions in order to use it in the coherent detection
of the transmitted signals. 
Opportunistic schedulers that are capable of exploiting the full MUD gains are based
on having continuously-updated channel state information (CSI) of all back-logged 
mobile users in the network at the central scheduler (i.e.~at the base station).
Thus, all mobile terminals inform the central scheduler about their CSI using
explicit feedback messages. 
As a result, a considerable portion of the air-link resources 
and a significant share of the battery energy of the mobile terminals are  
used for the CSI feedback instead of useful data traffic.
This fact has motivated many researchers to examine the feedback load of
opportunistic scheduling schemes\footnote{Similar to other papers in the 
literature such as \cite{YaAl:2006}, we 
refer to the systems that are based on full CSI feedback as multiuser selection diversity
(MUSelD) scheduling schemes.}
and to search for alternative schemes which can trade off some of the MUD gains
for considerable savings of the feedback load.
In \cite{ErOt:2007} and \cite{LoHeLaGeRaAn:2008}, extensive surveys on feedback 
reduction methods 
are provided. Note that the CSI feedback load is a common challenge in wireless
communication systems \cite{SpJSAC:2008}.
At present there is no general theory of single or multiuser wireless feedback
communication networks \cite{LoHeLaGeRaAn:2008}.
We can classify the solutions for the multiuser case into
two main approaches: (i) compression of the CSI messages by using
quantization methods or source coding techniques to exploit
the channel correlation across the air-link resource units, and (ii) reduction of the feedback
load by selectively choosing when to acquire a CSI feedback message based on
its likelihood to be useful in obtaining MUD gains.
The latter approach is generally more effective in reducing
the feedback load significantly and it is less complex to implement.

Under the theme of reduced-feedback opportunistic scheduling, Holter \textit{et~al.} proposed 
the multiuser ``switched-diversity'' (MUSwiD) scheduling scheme \cite{HoAlOiYa:2004}.
The basic principle in MUSwiD scheduling schemes is to find 
\textit{any acceptable user} (i.e.~ having good channel condition) 
instead of finding \textit{the best user} among all.
The term ``multiuser switched diversity'' was suggested in
\cite{HoAlOiYa:2004}, because the proposed scheduling scheme has
a similar principle of operation to the ``switch-based'' antenna
selection scheme used long-time ago in multiple-antenna receivers  
\cite{Shortall:1973}. 
It was suggested in \cite{HoAlOiYa:2004} to use a scheduling
strategy based on \emph{examining the CSI of the users sequentially instead of jointly}.
Once a ``good-channel'' user is found, the process of examining the channel conditions 
terminates, and that user is scheduled. The decision whether the channel condition of
a specific user is acceptable or not is assessed by a predefined threshold.
After the pioneer work \cite{HoAlOiYa:2004}, 
several modifications and enhancements have been proposed in the literature  
(e.g.~\cite{AlTeAl:2007}, \cite{NaKoAl:2008}, \cite{NaAl:2010_1} and \cite{NaAl:2010_2}).
The sate-of-the-art in this field are the recent works 
in \cite{NaAl:2010_1} and \cite{NaAl:2010_2} in which fundamental
concepts were suggested to enhance the performance of the MUSwiD schemes; namely the per-user 
thresholds \cite{NaAl:2010_1} and the post-user selection strategy \cite{NaAl:2010_2}.
In this paper, we basically build upon 
the per-user threshold approach adopted in \cite{NaAl:2010_1}.

The operation mode (i.e.~protocol) of the 
MUSwiD scheduling schemes \cite{NaAl:2010_1}  
is based on using a tiny-slotted feedback channel that 
is shared by all active users in the network.
The shared feedback channel was called the guard period in \cite{NaKoAl:2008}, \cite{NaAl:2010_1}.
Each mini time-slot of the shared feedback channel can be used 
to send a 1-bit flag signal\footnote{The time duration of the feedback channel is not long, and hence the MUSwiD scheduling scheme 
does not cause additional delay to the scheduling process.}. Furthermore, each mini-slot 
can be firmly accessed by a single user. The users are ordered into a sequence and 
assigned access to the mini-slots of the shared feedback channel accordingly. Per-user channel 
state thresholds are used. After a pilot signal is detected and a channel measurement 
is done, each user compares its current channel condition with respect to its 
associated channel threshold. A user sends a flag signal in its associated mini 
time-slot if it has above threshold channel condition, and all users before it 
in the feedback sequence have not sent flag signals. The first user 
to send a flag signal is the scheduled user to access the next resource unit. 
If the system adopts adaptive modulation and coding transmission \cite{SpProcAd:2007}, 
the selected user sends a full CSI message after the 1-bit flag signal in order
for the base station to adapt the transmission rate accordingly.

The feedback in MUSwiD systems is reduced significantly into only one feedback channel per resource unit
instead of per-user feedback channels due to the distributed scheduling
mechanism that makes the mobile terminals participate in the scheduling process
by comparing their channel condition locally against a pre-defined threshold, and
sending feedback flag signals using an ordered strategy which resolves contention. 
Another advantage of the system is that a user sends CSI feedback only ahead of
the resource units that it will be allocated instead of sending feedback for all resource
units, and this provides considerable savings in terms of battery life of mobile terminals.

Despite the evident feedback-reduction advantage of the state-of-the-art MUSwiD
schemes, there are some fundamental technical challenges that should be
addressed adequately before MUSwiD schemes can lend themselves for practical 
implementation. In our opinion, there are mainly three technical challenges:
\begin{itemize}
\item \textbf{Fairness:}
Maximizing the sum capacity is not always an appropriate
optimization criterion for realistic network scenarios since users usually have asymmetric
channel statistics. Furthermore, in MUSwiD schemes, the users' ordering strategy 
gives an advantage to the users who are placed in the first positions in the feedback sequence. 
It becomes likely that users placed in the latter positions of the sequence may not get channel access
despite having very strong channel. 
So, 
is it possible to achieve fairness in
MUSwiD schemes? and how? The current proposals in the topic (e.g.~\cite{NaAl:2010_2}, 
\cite{AlTeAl:2007}) suggest to keep changing the feedback sequence continuously 
in order to achieve fairness. We demonstrate in this paper that we 
can maintain fairness without this requirement.
\item \textbf{Centralized optimization:} As discussed in \cite{NaAl:2010_1}, the optimization 
of the feedback thresholds in MUSwiD systems is done at the central scheduler and it requires the 
knowledge of the statistics (i.e.~probability density functions (PDF))
of all users' channels. 
However, due to the CSI feedback reduction, the central scheduler
will not be able to have accurate estimates of the PDFs of the users' channels.
This will affect the optimality of the assigned per-user thresholds and will consequently
degrade the system performance. 
\item \textbf{Capacity-feedback tradeoff:} A comparison of MUSwiD schemes with full-feedback (MUSelD) 
opportunistic scheduling schemes is
needed to evaluate how much rate do we lose due to the feedback savings.
Such analysis is not provided in the available literature.
\end{itemize}

In this paper we provide a comprehensive study to answer the aforementioned
technical challenges.  
Furthermore, we aim in this work to persuade that MUSwiD scheduling systems
are actually attractive options for practical implementation in emerging
mobile broadband communication systems.
Toward this end, we take the following steps;
We provide detailed discussions to enhance our understanding about the attributes of the system
and how to optimize its performance. In particular, we characterize the achievable rate region
of MUSwiD systems.
Also, we show that the achievable rates in MUSwiD systems are comparable with 
selection-based systems although they are
significantly more economic in terms of CSI feedback load. 
Furthermore, we propose a novel MUSwiD scheduling scheme that achieves 
the proportional fairness criterion (\cite{Ke:1997}, \cite{ViTsLa:02}), which is preferable for practical 
implementation \cite{BeBlGrPaSiVi:2000}. 
We show that this can be achieved by proper per-user threshold optimization based on the objective function of maximizing
the sum of the logarithms of the achievable rates.
We demonstrate that our proposed scheme has a special interesting feature
that the solution of the corresponding optimization problem yields independent equations for each user, and hence
the threshold optimization can be decentralized, which overcomes the centralized optimization challenge.


The remainder of this paper is organized as follows. We provide in Section~\ref{sec:system_description} detailed
discussion about the achievable rates using MUSwiD scheduling schemes and
their optimization procedure.
We, then, provide in Section~\ref{sec:motivation} a motivation case study 
of the achievable rate region in a 2-user scenario. 
After that, we propose in Section~\ref{sec:proportional_fairness}
a novel proportional fair MUSwiD scheduling scheme and we discuss
its optimization procedure and demonstrate its practical advantages.
Next, we provide in Section~\ref{sec:performance_analysis} several numerical
examples to compare the performance of MUSwiD schemes with respect to
full-feedback MUSelD scheduling schemes.
Finally, we summarize the main conclusions in Section~\ref{sec:Conclusions}.

\section{Achievable Rates Using MUSwiD Systems}
\label{sec:system_description}

\subsection{System Model and General Assumptions}
\label{sec:system_model}

We consider the downlink\footnote{The proposed scheduling schemes can be
applied to the uplink as well based on the reciprocity of the uplink and downlink.
The receiver (i.e.~base station) transmits pilot signals
prior to every resource block, and the users (i.e.~mobile terminals) estimate
the uplink CSI from their measurements of the downlink channel condition.} 
in a single cell of a wireless communication system, and we consider
best-effort services so that delay constraints
are not taken into consideration in the scheduling decisions.
The base station communicates with the users through wireless block-fading channels.
We assume orthogonal access scheme in which the air-link resource units (i.e.~the channel blocks) 
are slotted in time and possibly in frequency as well. 
One user only can be scheduled per resource unit.
The time duration and the frequency bandwidth of one resource unit are assumed to be 
less than the coherence time and the coherence bandwidth of the fading channels so that
the channels can be modeled as constant additive white Gaussian noise (AWGN) channels within one resource unit
and varies randomly and independently from one resource unit to another. 
Furthermore, we assume that the base station transmits with constant power
over all resource units.

Assume that we have a number $M$ of active users in the network. The users are ordered
according to a strategy $\pi$ which is an injective (one-to-one) function.
User $i$ has the position $\pi(i)$ within the feedback sequence which defines the 
order by which the users can send flag signals to request being scheduled.
For simplicity of notation, we assume
that the users indices are consistent with their locations within the feedback sequence (i.e.~$\pi(i)=i$).
A user is scheduled if (i) its current channel condition is better than its associated 
channel threshold, and (ii) all users ahead of it in the feedback sequence  
have below-threshold channel condition.
In a mathematical context, the two conditions for a user $i$ to be scheduled\footnote{The scheduling
decision for a resource unit is based on the channel conditions of the users in this particular resource unit
only and it is independent of the channel conditions in other resource units. Thus, in multi-carrier systems,
the scheduling for each carrier (i.e.~frequency band) is done independently.} are:
\begin{subequations}\label{eq:conditions_for_scheduling}
\begin{align}
& r_i \geq r^*_i \label{eq:above_threshold_condition}\\
& \mathbf{r} \in \mathcal{S}_i, \label{eq:no_flag_condition}
\end{align}
\end{subequations} 
where $r_i$ is the achievable rate by user $i$, $r^*_i$ is the 
channel threshold associated with user $i$, and $\mathbf{r}=[r_1 \; r_2 \cdots r_M]$ is the vector of
acheivable rates of all users. The per-user thresholds
were presented in terms of SNR in \cite{NaAl:2010_1}.
However, we prefer in this paper to present the thresholds in terms of achievable
rates (i.e.~channel capacity) because it is a more generic framework and enables extending the results
into multiple antenna scenarios.
The event $\mathcal{S}_i$ in \eqref{eq:no_flag_condition} is defined as:
\begin{equation}
\label{eq:S_i_event_definition}
\mathcal{S}_i \doteq \left\{\mathbf{r} \in \mathcal{R}_+^M\; | \;  r_j<r^*_j \quad \forall j<i \right\}.
\end{equation}

Note that $\mathcal{S}_1=\mathcal{R}_+^M$ and that $\mathcal{S}_M \subset \mathcal{S}_{M-1} \subset \cdots \subset \mathcal{S}_1$.
As an alternative mathematical representation, we can combine the two conditions in \eqref{eq:conditions_for_scheduling}
into one expression $\mathbf{r} \in \mathcal{S}_i \setminus \mathcal{S}_{i+1}$.
We assume that the fading processes of the users' channels are 
stationary, independent of each other and have continuous PDF of the achievable rates
($f_{R}(r)$). As such, we can write
$f_{R_1\cdots R_M}(r_1,\cdots,r_M)=\prod^M_{i=1}f_{R_i}(r_i)$.

In the numerical examples in the paper, we assume that the base station 
and the users' terminals are equipped with a single antenna to transmit/receive
and thus the relation of the achievable rate -- i.e.~capacity -- (denoted by $r$) and the SNR 
(denoted by $\gamma$) is given by the classical capacity relation of AWGN channels 
$r_i=\log(1+\gamma_i)$.
We can show using simple steps 
that the PDF of achievable rate $f_R(r)$ can be expressed in terms of the PDF
of SNR $f_\Gamma(\gamma)$ in this case as
$f_R(r)=\exp(r)\; . \; f_\Gamma\left(\exp(r)-1\right)$.

The extension to multiple antennas is straightforward as long as
a single user only is served per resource unit. The per-user thresholds can still be 
presented using a single value 
instead of multiple SNR thresholds for every transmit/receive antenna pair.
The appropriate capacity formulas  
should be used in the derivation of the PDF of the achievable rates
in this case.

\subsection{Statistical Analysis of the Users Expected Achievable Rates}
\label{sec:performance_equations}

The long-term expected (i.e.~average) achievable rate by each user in MUSwiD systems
was analyzed in \cite{NaAl:2010_1} in terms of $f_{\Gamma_{i}}(\gamma_i)$. 
In this section, we briefly review these results in a variant representation using 
$f_{R_{i}}(r_i)$. 
We denote the conditional expected value of the achievable rate by user $i$ given that
\eqref{eq:no_flag_condition} is satisfied as $R^c_i$ defined mathematically as
\begin{equation}
\label{eq:conditional_rate}
R^c_i \doteq E[r_i|\mathbf{r} \in \mathcal{S}_i]=\int^\infty_{r^*_{i}}r f_{R_{i}}(r) dr,
\end{equation}
where $E[\;]$ is the expectation operator.
Since we assume that the fading channels of the users are independent, the
event $\mathbf{r} \in \mathcal{S}_i$ happens with probability
$\Pr\{\mathbf{r} \in \mathcal{S}_i\}=\prod_{j<i}F_{R_j}(r^*_j)$.
Furthermore, the unconditional
expected value of the achievable rate by user $i$, denoted as $R_i$, equals
\begin{equation}
\label{eq:expected_achievable_rate}
R_i \doteq E[r_{i}]=E[r_i|\mathbf{r} \in \mathcal{S}_i] \; . \; \Pr\{\mathbf{r} \in \mathcal{S}_i\}=\int^\infty_{r^*_{i}}r f_{R_{{i}}}(r) dr \; . \; \prod_{j<i}F_{R_j}(r^*_j)
\end{equation}

Similarly, the expected percentage of resource units scheduled to user $i$
(i.e.~channel access ratio $\text{AR}_i$) is given as
$\text{AR}_i= \left(1-F_{R_i}(r^*_i)\right)  \; . \; \prod_{j<i}F_{R_j}(r^*_j)$.
The expected achievable rates in single-input-single-output (SISO) channels can be presented equivalently in terms of SNR threshold values
($\gamma^*_j$) as \cite{NaAl:2010_1}:
\begin{equation}
R_i= \int_{\gamma^*_i}^\infty f_{\Gamma_i}(\gamma)  \log(1+\gamma) d\gamma  \; . \; 
\prod_{j<i}F_{\Gamma_j}(\gamma^*_j).
\end{equation}

\subsection{Per-User Thresholds Optimization}
\label{sec:optimization}

From \eqref{eq:expected_achievable_rate}, it is clear
that the system performance is dependent on (i) the chosen strategy to order the users in the feedback sequence
and (ii) the channel thresholds $r^*$ of all users. The channel threshold of one
user does not only affect its achievable rate alone, but additionally all other users
placed next in the feedback sequence. 
In this Section we discuss the joint optimization of the per-user thresholds. 
However, we assume first that the feedback sequence is fixed beforehand.
We discuss in later sections the selection of the feedback sequence.
In \cite{NaAl:2010_1}, the optimization of the per-user thresholds was derived
with the objective of maximizing the aggregate (sum) capacity (achievable rate) of all
users in the network. We first summarize these results 
in the context of this paper
based on representing the per-user thresholds in terms of achievable rates.
We then provide a generalized framework to optimize the per-user thresholds taking
fairness into consideration.

The optimal per-user thresholds are obtained by solving the following optimization problem:
\begin{equation}
\label{eq:optimization_basic}
\left\{\hat{r^{*}_1}, \cdots ,\hat{r^{*}_M}\right\} = \arg 
\max_{\left\{r^*_1, \cdots ,r^*_M\right\}} \Phi,
\end{equation}
where we use the notation $\hat{r^{*}_i}$ to denote the optimal value
for the threshold $r^*_i$ under the objective function $\Phi$. In the special case of
maximizing the sum achievable rate, $\Phi$ is defined as
\begin{equation}
\label{eq:objective_sumC}
\Phi=\sum_{i=1}^M R_i,
\end{equation}
where the expected achievable rates $R_i$ follow \eqref{eq:expected_achievable_rate}.
In order to solve \eqref{eq:optimization_basic} with \eqref{eq:objective_sumC}, we search 
at the points at which the gradient equals zero (i.e.~the stationary points): 
\begin{equation}
\label{eq:gradient_basic}
\frac{\partial \Phi}{\partial r^*_i} = 
  0, \quad \forall i\leq M.
\end{equation}

The derivative $\frac{\partial R_j}{\partial r^*_i}$ is obtained as follows:
\begin{equation}
\label{eq:derivative}
\frac{\partial R_j}{\partial r^*_i} = 
\left\{
   \begin{array}{ll}
        0  & : \quad i > j\\
        -r^*_i \; f_{R_i}(r^*_i) \; \prod_{k<i}F_{R_k}(r^*_k) & : \quad i = j \\
	\frac{R_j}{F_{R_i}(r^*_i)}\;f_{R_i}(r^*_i) & : \quad i < j
    \end{array} 
 \right. 
\end{equation}

The derivative of $R_i$ with respect to $r^*_i$ (second line in \eqref{eq:derivative})
is obtained by applying the first fundamental theorem of calculus (e.g.~\cite{Ap:1967}).
We can alternatively write $\frac{\partial R_j}{\partial r^*_i} \; :i<j$ as:
\begin{equation}
\label{eq:alternative_representation}
\frac{\partial R_j}{\partial r^*_i} =
R^c_j \; f_{R_i}(r^*_i) \; \prod_{k<j, \; k\neq i}F_{R_k}(r^*_k) \quad : \quad i < j.
\end{equation}

By inserting \eqref{eq:derivative} into \eqref{eq:gradient_basic}, we obtain
$\hat{r^{*}_i} \; f_{R_i}(\hat{r^{*}_i}) \; \prod_{k<i}F_{R_k}(\hat{r^{*}_k}) =  
\frac{f_{R_i}(\hat{r^{*}_i})}{F_{R_i}(\hat{r^{*}_i})}\;\sum^M_{j>i}R_j$,
yielding 
\begin{equation}
\label{eq:threshold_formula_1}
\hat{r^{*}_i}  =  \frac{\sum^M_{j>i}R_j}{\prod_{k<i+1}F_{R_k}(\hat{r^{*}_k})},
\end{equation}
where the assumptions ($r^*_i \neq 0, \; \forall i<M$)
and ($f_{R_i}(r)\neq 0, \; :r>0$) are used. We can re-write \eqref{eq:threshold_formula_1} as
\begin{equation}
\label{eq:threshold_formula_2}
\hat{r^{*}_i} = E\left[\Phi|\mathbf{r} \in \mathcal{S}_{i+1} 
\text{ and } 
r^*_j=\hat{r^*_j} \; \; \forall j>i\right] .
\end{equation}

From \eqref{eq:threshold_formula_2} and using a simple intuitive explanation, we can describe
the basic principle for optimizing the per-user thresholds in switched diversity
systems as trying to maximize the outcome (which is the achievable rate in our case) of a random experiment
(which is examining the channel condition, i.e.~achievable rate, of one user in our case) with the possibility to repeat the experiment 
up to a limited number of trials (which is the total number of users in our case). 
After the experiment is executed once and its output is observed, 
we can either choose to accept its outcome and stop repeating the experiment, 
or opt to repeat the random experiment taking into consideration that we will lose the output that is
already observed and the expected output of the new trial of the experiment will be totally independent
of the previous ones\footnote{Note that \eqref{eq:threshold_formula_2} is invalid 
in case of post-user selection \cite{NaAl:2010_2}
since the last trial (i.e.~post-user selection) is dependent on another trial (the one related
to the post-selected user). Thus, the optimization of MUSwiD systems with post-user selection is not
as straightforward as in the case of MUSwiD systems without post-user selection.}.
As an intuitive guideline to the decision making of choosing whether to repeat the experiment or 
to accept the observed output (which corresponds to the decision to send a flag signal by the corresponding user in our case), 
we will decide not to repeat the experiment if the observed output
is very good so that we do not expect to obtain such a good output if we repeat the experiment. Similarly,
we will decide to repeat the experiment if the observed output is low so that we expect that most likely we
will obtain a better result by repeating the experiment.  
The optimal solution to this decision making problem is that we compare the
observed output with the expected value for the outcome of the allowed number of trials to repeat the experiment. 
If the current outcome is higher than the expected value for repeating the experiment,
we accept it and stop repeating the experiment and vice versa. 

We can write \eqref{eq:threshold_formula_1} alternatively as
\begin{equation}
\label{eq:thresholds_formula_3}
\hat{r^{*}_i}=\sum_{j>i} \left[R^c_j \; \prod_{i<k<j}F_{R_k}(\hat{r^{*}_k}) \right].
\end{equation}

We can see that the optimal threshold of each user depends on the optimal thresholds of 
all users that are placed after it according to the feedback sequence.
Thus, the per-user thresholds can be obtained using a backward successive approach
starting from the last user in the sequence.
Note that it is intuitive to predict that the threshold of the last user 
in the sequence is zero since we do not apply 
a post-user selection strategy \cite{NaAl:2010_2} or power control. 
\begin{equation}
\label{eq:threshold_M}
\hat{r^{*}_M}=0
\end{equation}

Furthermore, by using some mathematical manipulations \cite{NaAl:2010_1},
we can use the following formula for the backward successive approach
for obtaining the per-user thresholds:
\begin{equation}
\label{eq:thresholds_formula_4}
\hat{r^{*}_i}=\int^\infty_{\hat{r^{*}_{i+1}}}r f_{R_{i+1}}(r) dr 
\; + \; \hat{r^{*}_{i+1}} \; F_{R_{i+1}}\left(\hat{r^{*}_{i+1}}\right).
\end{equation}

Also, we can show by simple mathematical manipulations similar to \eqref{eq:threshold_formula_2}
and \eqref{eq:thresholds_formula_4} that the maximum achievable sum rate is:
\begin{equation}
\label{eq:max_sum_rate}
\max \Phi = \int^\infty_{\hat{r^{*}_{1}}}r f_{R_{1}}(r) dr 
\; + \; \hat{r^{*}_{1}} \; F_{R_{1}}\left(\hat{r^{*}_{1}}\right),
\end{equation}
where $\Phi$ is the sum achievable rate \eqref{eq:objective_sumC}.

As well-known, maximizing the sum achievable rate is not always a suitable
optimization criterion for multiuser networks since it creates fairness problems. 
Motivated by this fact, we provide here a more generic framework to optimize the performance of MUSwiD systems.
From an information-theoretic point of view (e.g.~\cite{TsHa:98}, \cite{LiGo1:2001}),
the objective in multiuser channels 
is to operate at the boundary surface of the achievable-rate region. The points 
on the boundary surface are Pareto-optimal, which means that we cannot increase the 
achievable rate of one user without decreasing the achievable rate of another user. 
The objective of scheduling schemes in multiuser networks should be to achieve 
Pareto-optimality\footnote{As discussed in \cite{ShGo3:2008},
there is no contradiction between the two objectives of (i) efficient resource
allocation by designing scheduling schemes leading to operating at the points 
on the boundary surface of the achievable rate region, and (ii) achieving 
fairness among the users as well as maintaining the QoS requirements, which can be done
by controlling the operating point of the system based on proper selection 
of $\pmb{\mu}$ (the vector of the users' weighting factors).
The specific selection of $\pmb{\mu}$ to meet fairness requirements or QoS constraints is a different topic that is not specific to this work on MUSwiD schedulers.
Few examples of the many possible approaches suggested in the literature to select the specific operating 
point of the system are (i) the fairness-based approach, such as 
the proportional fairness scheduler \cite{ViTsLa:02} and 
the flexible resource-sharing constraints scheduler \cite{ShGo4:2008}, 
(ii) the utility-maximization-based approach \cite{SoLi3:2005}, and
(iii) the QoS constraints based approach \cite{WaGiMa:2007}.}.
The points on the boundary surface of the achievable rate region are obtained by 
maximizing a weighted sum of the rates. By varying the weights we can scan all 
points on the boundary surface.
Thus, we propose to use a weighted sum of the achievable rate as the 
optimization objective for \eqref{eq:optimization_basic}:
\begin{equation}
\label{eq:objective_weighted_sum}
\Phi=\sum_{i=1}^M \mu_i \; R_i.
\end{equation}

Note that another common approach is to maximize the sum of concave and monotonically increasing utility functions
of the achievable rates of the users $\Phi=\sum_{i=1}^M U(R_i)$. 
As discussed in \cite{WaGiMa:2007}, this is interlinked with the objective of maximizing a wighted
sum of the rates by using $\mu_i=U'(R_i)$. 

By repeating the same procedure used for maximizing the sum achievable rates case,
we obtain the following results for optimizing the per-user thresholds with the objective of maximizing 
a weighted sum of the achievable rates. Equations \eqref{eq:threshold_formula_1}, \eqref{eq:thresholds_formula_3} 
and \eqref{eq:thresholds_formula_4} are replaced
by \eqref{eq:threshold_mu_formula_1}, \eqref{eq:thresholds_mu_formula_3} and \eqref{eq:thresholds_mu_formula_4} respectively. 

\begin{equation}
\label{eq:threshold_mu_formula_1}
\mu_i \; \hat{r^{*}_i} =  \frac{\sum^M_{j>i} \mu_j \; R_j}{\prod_{k<i+1}F_{R_k}(\hat{r^{*}_k})},
\end{equation}

\begin{equation}
\label{eq:thresholds_mu_formula_3}
\mu_i \; \hat{r^{*}_i}=\sum_{j>i} \left[\mu_j \; R^c_j \; \prod_{i<k<j}F_{R_k}(\hat{r^{*}_k}) \right],
\end{equation}

\begin{equation}
\label{eq:thresholds_mu_formula_4}
\mu_i \; \hat{r^{*}_i}=\mu_{i+1} \left[\int^\infty_{\hat{r^{*}_{i+1}}}r f_{R_{i+1}}(r) dr 
\; + \; \hat{r^{*}_{i+1}} \; F_{R_{i+1}}\left(\hat{r^{*}_{i+1}}\right)\right].
\end{equation}

Note that \eqref{eq:threshold_M} is also valid for the generic case of
maximizing a weighted sum of the achievable rates \eqref{eq:objective_weighted_sum}. 
Equation \eqref{eq:threshold_formula_2} is replaced by:
\begin{equation}
\label{eq:threshold_mu_formula_2}
\mu_i \; \hat{r^{*}_i} = E\left[\Phi|\mathbf{r} \in \mathcal{S}_{i+1}
\text{ and } 
r^*_j=\hat{r^*_j} \; \; \forall j>i\right].
\end{equation}

The maximum weighted sum of the achievable rates can be expressed as
\begin{equation}
\label{eq:max_weighted_rate}
\max \Phi = \mu_1 \left[\int^\infty_{\hat{r^{*}_{1}}}r f_{R_{1}}(r) dr 
\; + \; \hat{r^{*}_{1}} \; F_{R_{1}}\left(\hat{r^{*}_{1}}\right)\right],
\end{equation}
where $\Phi$ is defined in \eqref{eq:objective_weighted_sum}.

For SISO channels, the optimal per-user thresholds in terms of SNR are computed according to
\begin{equation}
\mu_i\log(1+\hat{\gamma^{*}_i}) = \mu_{i+1} \left[ \int_{\hat{\gamma^{*}_{i+1}}}^\infty f_{\Gamma_{i+1}}(\gamma) 
 \log(1+\gamma) d\gamma
+\log(1+\hat{\gamma^{*}_{i+1}})F_{\Gamma_{i+1}}(\hat{\gamma^{*}_{i+1}})\right],
\end{equation}
which is done in a backward successive approach starting with $\hat{\gamma^*_M}=0$.

In the numerical examples used in this paper, we assume SISO Rayleigh block-faded channels.
We show in Table~\ref{table:SISO_Rayleigh} the closed-form formulas to characterize
the performance of the system and the optimization of the thresholds.
In order to obtain simple closed-form expressions, the formulas are presented in terms of 
the SNR-based thresholds $\gamma^*_i$.

\begin{table*}[P] 
\caption{Multiuser Switched Diversity System - SISO Rayleigh Block-Fading}
\label{table:SISO_Rayleigh}
\centering
\HRule \\
Expected achievable rates:
\begin{equation}
R_i= \left[\exp\left(\frac{-\gamma^*_i}{\bar \gamma_i}\right)\log(1+\gamma^*_i)
+\exp\left(\frac{1}{\bar \gamma_i}\right)E_1\left(\frac{1+\gamma^*_i}{\bar \gamma_i}\right)\right]
 \; . \; \prod_{j<i}\left(1-\exp\left(\frac{-\gamma^*_j}{\bar \gamma_j}\right)\right),
\end{equation}
where $\bar \gamma_i$ is the average SNR of user $i$ and $E_1$ is the exponential integral function:
\begin{equation}
\label{eq:E1_definition}
E_1(x)\equiv\int^{\infty}_{x}\frac{\exp(-u)}{u}du .
\end{equation}

Expected access ratio:
\begin{equation}
\text{AR}_i= \exp\left(\frac{-\gamma^*_i}{\bar \gamma_i}\right)
 \; . \; \prod_{j<i}\left(1-\exp\left(\frac{-\gamma^*_j}{\bar \gamma_j}\right)\right) .
\end{equation}

The optimal feedback thresholds are computed in a backward successive approach,
starting from $\hat{\gamma^{*}_M}=0$, according to:
\begin{equation}
\label{eq:thresholds_mu_SNR}
\mu_i\log(1+\hat{\gamma^{*}_i})= 
\mu_{i+1} \left[\exp\left(\frac{1}{\bar \gamma_{i+1}}\right)E_1\left(\frac{1+\hat{\gamma^{*}_{i+1}}}{\bar \gamma_{i+1}}\right)
+\log(1+\hat{\gamma^{*}_{i+1}})\right] .
\end{equation}

In the special case of maximizing the sum achievable rates, all weighting factors $\mu_i$ in \eqref{eq:thresholds_mu_SNR} are equal,
and the maximum sum achievable rate can be expressed as: 
\begin{equation}
\label{eq:sum_capacity_iid_MUSwiD}
\max \sum_{i=1}^M R_i=\exp\left(\frac{1}{\bar \gamma_1}\right)E_1\left(\frac{1+\hat{\gamma^{*}_1}}{\bar \gamma_1}\right)
+\log(1+\hat{\gamma^{*}_1}) .
\end{equation}

\HRule \\
\end{table*}

\section{Motivation Case Study -- Achievable Rate Region in 2 user Scenario}
\label{sec:motivation}

Studying the achievable rate region in 2-user scenario is a useful tool in order to 
get basic insights regarding the performance limits of the system and the tradeoff between
maximizing the sum capacity and maintaining fairness among the users.
In order to characterize the achievable rate region in MUSwiD schemes,
we solve \eqref{eq:optimization_basic} with the objective 
function \eqref{eq:objective_weighted_sum} for different values of the weighting factors $\mu$,
ranging from ($\mu_1=1$, $\mu_2=0$) to ($\mu_1=0$, $\mu_2=1$).

\begin{figure}[P] 
 \begin{center}
\includegraphics[scale=0.7]{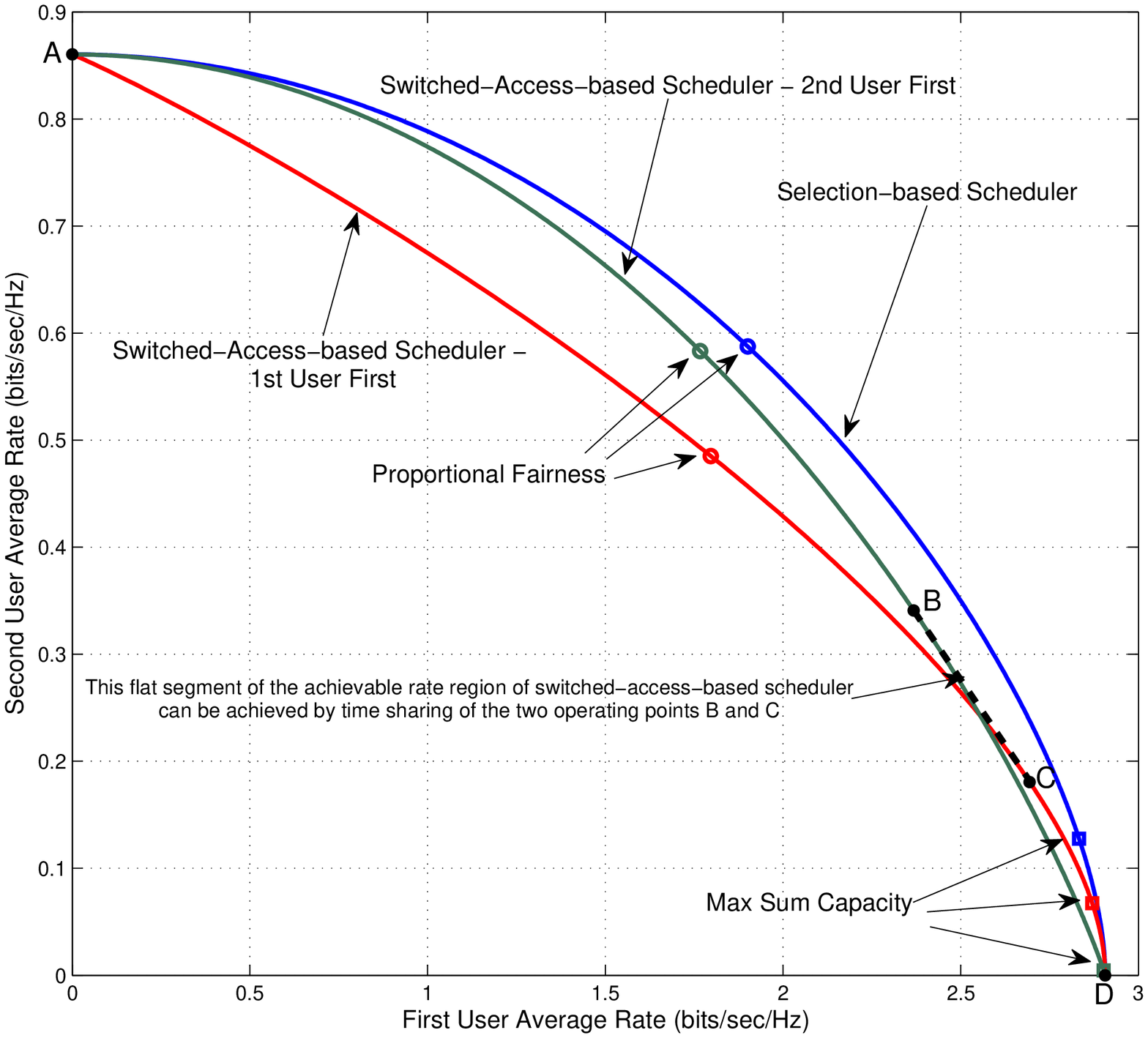}
\caption{Achievable rate region of selection-based and switched-access-based
diversity systems in a two-user scenario. The channels are Rayleigh block-faded with
average SNR of 10 dB and 0 dB for the first user and second user, respectively.}
   \label{fig:2D_rate_region}
 \end{center}
\end{figure}

In the numerical example of 2-user scenario shown in Fig.~\ref{fig:2D_rate_region}, we 
assume that both users as well as the base station are equipped with single antennas.
Furthermore, we assume that both users have Rayleigh block-fading channels but with different
expected average values. Table~\ref{table:SISO_Rayleigh} summarizes the main
formulas under these particular assumptions.
We show in Fig.~\ref{fig:2D_rate_region} the achievable rate region for the two possible feedback
sequences.
In the first case, the user with better
average SNR is placed in the first position of the sequence.
While in the second case, the user with lower average SNR
is placed first.  
Furthermore, we compare with the achievable rate region of the full feedback selection-based MUSelD scheme,
which is known from the literature (e.g.~\cite{ShGoTh_eusipco:2009}, \cite{ShGo3:2008}).
A summary of the formulas to characterize the achievable rates in MUSelD schemes is provided 
in Section~\ref{sec:recap_selection}.
We show in Fig.~\ref{fig:2D_rate_region} some special cases including the maximum sum rate and the proportional
fairness operating points. Detailed discussion about proportional fair MUSwiD scheduling
is provided in Section~\ref{sec:proportional_fairness}. 
The main conclusions obtained from this motivation case study carry over to the general case of $M$ users.
We summarize below the key learnt messages.

The achievable rates by MUSwiD scheduling schemes are always close to the 
achievable rates with full feedback MUSelD scheduling schemes. The little loss
in the achievable rates is an acceptable trade-off for evident reductions in the CSI feedback load.
Also, the maximum sum rate is an unfair operating point in MUSelD scheme as well as in
MUSwiD schemes. Changing the feedback sequence in MUSwiD systems, while optimizing the thresholds to maximize 
the sum rate, does not solve the fairness issues. Furthermore, unlike the common belief 
in early works in MUSwiD schemes such as \cite{NaAl:2010_2} 
and \cite{AlTeAl:2007}, Fig.~\ref{fig:2D_rate_region} demonstrates that we can actually
achieve fairness in MUSwiD schemes without the need of 
alternating between feedback sequences. However, the per-user
channel thresholds should be adjusted properly to allow achieving fairness. Also, we can achieve
fairness regardless of the used feedback sequence. 

We observe also that alternating between feedback sequences can in some cases 
(the line $BC$ in Fig.~\ref{fig:2D_rate_region}) be the optimal solution. However, the optimization of
a MUSwiD scheduler including alternating between the feedback sequences as 
a degree of freedom is not simple and require complex algorithms
with significant computation load in order to find the optimal per-user thresholds
for each used sequence as well as the average time percentage of each 
used sequence since some sequences may need to be used more frequently than others.
Furthermore, the real-time implementation of MUSwiD schedulers with the option of
mixing up between different feedback sequences 
adds more control messages communication since the base station
should inform the users about all used feedback sequences and their associated per-user
thresholds.
On the other hand, Fig.~\ref{fig:2D_rate_region} demonstrates that it is almost sufficient to 
use one sequence to operate on or close to the achievable rate region limits.
Furthermore, it provides a practical scheduler design with low computation complexity
and feasible implementation procedure.
The loss in terms of performance will be void for most operating points
and negligible for some ranges of Pareto-optimal operating points.

Finally, we observe that choosing the proper feedback sequence is important in order to operate
at the boundary of the achievable rate region. However, for a number $M$ of users, we have a number $M!$
of possible feedback sequences. Thus, even for a relatively small number of users, comparing the performance
for all possible feedback sequences in order to find the optimal one is computationally expensive.
To simplify this task, we propose instead a very simple rule  
based on Fig.~\ref{fig:2D_rate_region} and the numerical results in Section~\ref{sec:numerical_results}. This rule is that
when the objective is to maximize the sum achievable rate in the network,
we should use a feedback sequence 
in which the users are sorted in descending order of their expected channel condition.
On the other hand, when fairness is taken into consideration,
we should use a feedback sequence in which the users are sorted in
ascending order of their expected channel condition.

\section{Proposed Scheme -- Proportional Fair Scheduler}
\label{sec:proportional_fairness}

Proportional fairness \cite{Ke:1997} is a well-known fairness criterion that
provides a good trade-off between the aggregate rate over the network and
fairness among users. 
Proportional fairness resolves this conflict by allocating to each user a transmission rate
relative to its channel condition without affecting the rates of other users.
Proportional fairness was suggested for full-feedback MUSelD scheduling schemes
in \cite{ViTsLa:02}, and it was applied in industry such as in the IS-856 
standard \cite{BeBlGrPaSiVi:2000}. In this paper, we propose to apply proportional 
fairness into MUSwiD scheduling schemes.

The optimization objective function $\Phi$ in case of proportional fairness is to maximize the 
product of the expected achievable rates of the users $\prod_{i=1}^M R_i$, or equivalently, to maximize
the sum of the logarithms of the expected achievable rates:
\begin{equation}
\label{eq:objective_PF}
\Phi=\sum_{i=1}^M \log\left(R_i\right).
\end{equation}

In order to optimize the per-user thresholds to achieve proportional fairness,
we solve \eqref{eq:optimization_basic} with the 
objective function \eqref{eq:objective_PF}. We find the points at which the gradient
equals zero, yielding
\begin{equation}
\label{eq:gradient_PF}
\frac{\partial \Phi}{\partial r^*_i} = 
\sum_{j=1}^M\frac{\partial \log(R_j)}{\partial r^*_i} =
\sum^M_{j=1} \frac{\frac{\partial R_j}{\partial r^*_i}}{R_j} =
  0, \quad \forall i\leq M,
\end{equation}
where $\frac{\partial R_j}{\partial r^*_i}$ is obtained in \eqref{eq:derivative}.
By solving \eqref{eq:gradient_PF} we obtain:
\begin{equation}
\label{eq:threshold_PF_1}
\frac{\hat{r^{*}_i} \; f_{R_i}(\hat{r^{*}_i}) \; \prod_{k<i}F_{R_k}(\hat{r^{*}_k})}
{R^c_i \; \prod_{k<i}F_{R_k}(\hat{r^{*}_k})} =  
\sum^M_{j>i} \frac{f_{R_i}(\hat{r^{*}_i})}{F_{R_i}(\hat{r^{*}_i})} .
\end{equation}

We can simplify \eqref{eq:threshold_PF_1} as
$\frac{\hat{r^{*}_i} \; f_{R_i}(\hat{r^{*}_i})}{R^c_i}  \; \frac{\prod_{k<i}F_{R_k}(\hat{r^{*}_k})}
{\prod_{k<i}F_{R_k}(\hat{r^{*}_k})}   
=\frac{f_{R_i}(\hat{r^{*}_i})}{F_{R_i}(\hat{r^{*}_i})} \left(M-i\right)$.
With the assumptions that ($r^*_i \neq 0, \; \forall i<M$)
and ($f_{R_i}(r)\neq 0, \; :r>0$) and by substituting for $R^c_i$ using \eqref{eq:conditional_rate} we obtain
\begin{equation}
\label{eq:PF_main}
\frac{\hat{r^{*}_{i}} \; F_{R_{i}}(\hat{r^{*}_{i}})}{\int^\infty_{\hat{r^{*}_{i}}}r f_{R_{i}}(r) dr}=M-i
\end{equation}

We can observe from \eqref{eq:PF_main} that the optimization of the proportional fair 
scheduler has a very interesting and unique feature.
The optimal achievable rate threshold of any user is only dependent on its channel statistics alone
and its location (index) within the feedback sequence.
Thus, we can optimize the system by solving $M$ independent equations instead of solving 
dependent equations successively as in the general case of MUSwiD scheduling schemes which 
was discussed in Section~\ref{sec:optimization}.
Among all Pareto-optimal operating points, the independent equations feature
is uniquely valid in the case of the proportional fair operating point.
This feature has a significant advantage from practical implantation perspective 
because it enables every user to obtain its optimal threshold value locally.
This overcomes the technical challenge of centralized threshold optimization of 
conventional MUSwiD schemes since every user can have
accurate prediction of its channel statistics while the base station cannot have such accurate measures
of the PDFs of the users' channels without explicit feedback from all users.
This feature is compatible with the main theme of MUSwiD schemes, which is to limit the feedback load.

Note that the optimization of the proportional fair scheduler \eqref{eq:PF_main} 
is consistent with the optimization procedure of the generic scheduling criteria of maximizing a weighted sum
of the achievable rates discussed in Section~\ref{sec:optimization}. 
In the case of proportional fairness, the weighting factor of
each user is inversely proportional with its expected achievable rate \cite{ViTsLa:02}.
By substituting
$\mu^{\text{PF}}_i=\frac{1}{R_i}$
into \eqref{eq:threshold_mu_formula_1} we obtain \eqref{eq:PF_main}.

In the case of SISO channels, We can alternatively present \eqref{eq:PF_main} 
in terms of SNR thresholds as:
\begin{equation}
\label{eq:PF_main_SNR}
\frac{\log(1+\hat{\gamma^{*}_{i}}) \; F_{\Gamma_{i}}(\hat{\gamma^{*}_{i}})}
{\int^\infty_{\hat{\gamma^{*}_{i}}} f_{\Gamma_{i}}(\gamma) \; \log(1+\gamma) \; d\gamma}=M-i
\end{equation}

\begin{figure}[htb]
 \begin{center}
\includegraphics[scale=0.7]{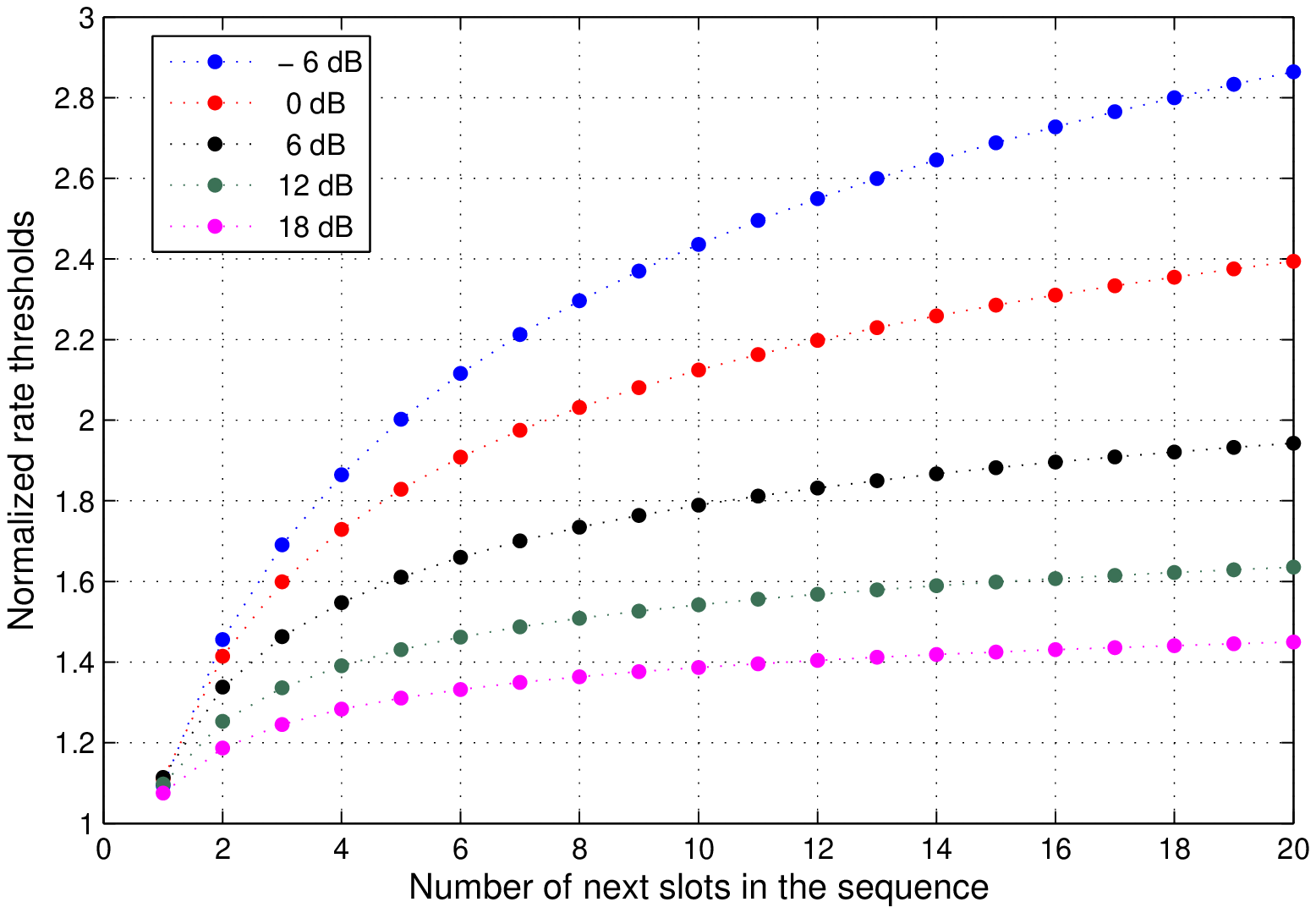}
\caption{Normalized achievable rate thresholds to achieve proportional fairness 
for SISO Rayleigh block-fading conditions with
different values for the average SNR plotted versus the number of next users
in the sequence.}
   \label{fig:R_thr_PF}
 \end{center}
\end{figure}

\begin{figure}[htb]
 \begin{center}
\includegraphics[scale=0.7]{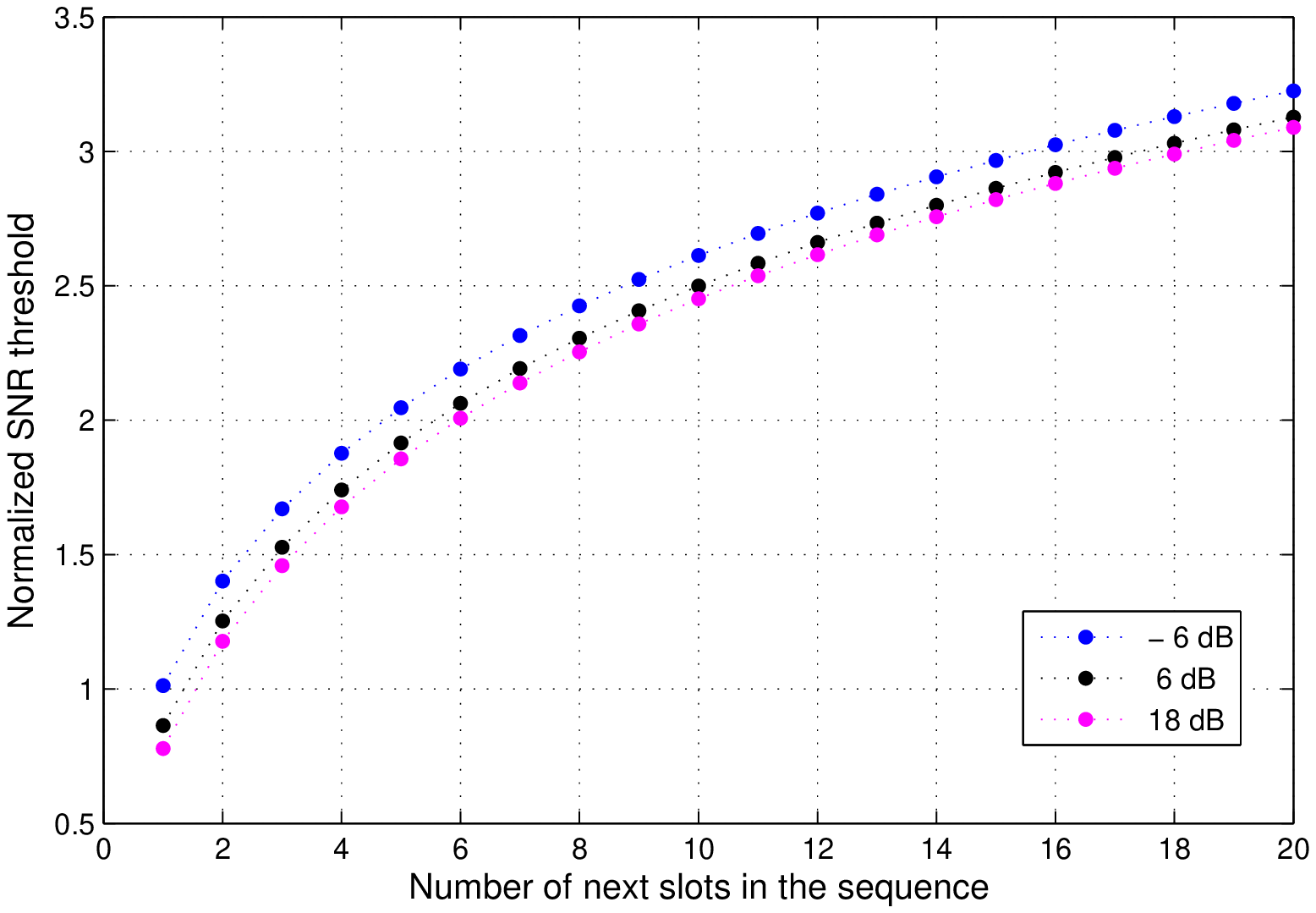}
\caption{Normalized SNR thresholds to achieve proportional fairness for SISO Rayleigh block-fading conditions with
different values for the average SNR plotted versus the number of next users
in the sequence.}
   \label{fig:SNR_thr_PF}
 \end{center}
\end{figure}

The left hand side of equation \eqref{eq:PF_main} is a monotonically increasing function of 
$\hat{r^{*}_{i}}$. Thus, the solution of \eqref{eq:PF_main} always exists and it is unique.
The solution can be obtained using simple numerical methods such as the bisection method.
Alternatively, the results can be obtained for standard channel models and 
stored in the mobile terminals using look-up tables versus the user index within the feedback sequence.
Fig.~\ref{fig:R_thr_PF} and Fig.~\ref{fig:SNR_thr_PF} show the optimal per-users thresholds
in the proportional fair scheduler in terms of achievable rates and SNR respectively
for SISO Rayleigh block-faded channels.
The per-user thresholds in both figures are normalized with respect to the average achievable rate
and average SNR respectively.

\begin{figure}[htb]
 \begin{center}
 \includegraphics[scale=0.88]{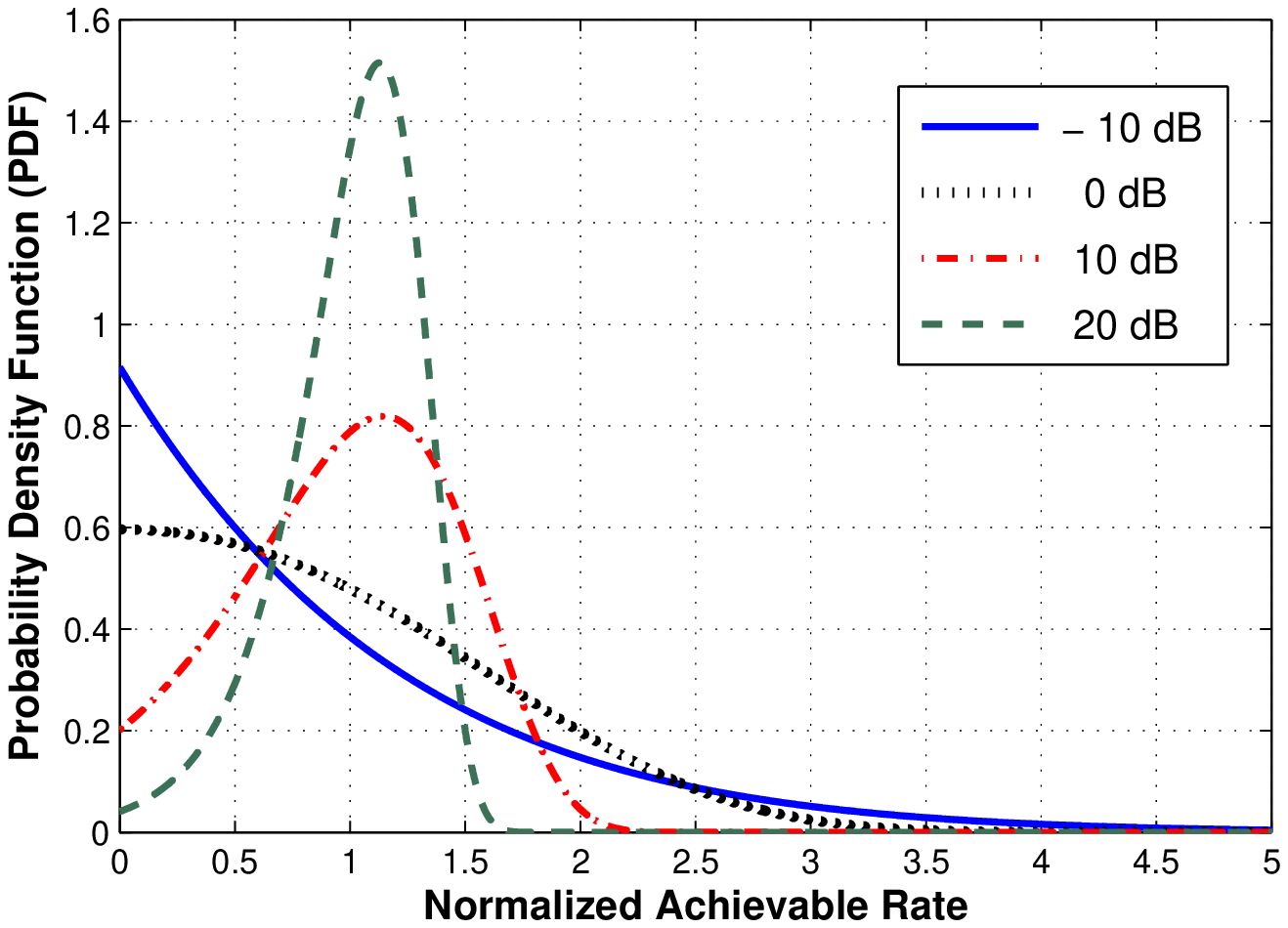}
\caption{The PDF of the normalized achievable rates over Rayleigh block-fading channels.
The achievable rates are normalized with respect to the average achievable rate
over the channel.}
   \label{fig:normalized_rate_PDF}
 \end{center}
\end{figure}

We observe from Fig.~\ref{fig:R_thr_PF} and Fig.~\ref{fig:SNR_thr_PF} that
as the number of next users in the sequence increases (meaning being placed in the first positions in the sequence),
the corresponding per-user threshold increases.
Thus, the users in the first places of the sequence are requested to achieve high rates (MUD gains)
with low expected success ratio, while the users at the last places is expected to achieve lower
rate gains but with higher success ratio. 
It is intuitive to predict that placing the users who have wider dynamic range of channel
variations in the first positions of the sequence is advantageous because these users
are more capable of achieving high rate gains when their channel condition is at its peak.
Thus, it is better in the feedback sequence to sort the users in ascending order of their expected (average) SNR.
This is due to the fact that at low SNR, the achievable rate formula $r=\log(1+\gamma)$
becomes almost linear and consequently more sensitive to the variations in SNR. 
Fig.~\ref{fig:normalized_rate_PDF} shows the PDF of the normalized achievable rates
for SISO Rayleigh block-fading channels with different average values.
We can see from Fig.~\ref{fig:normalized_rate_PDF} that at high average SNR ($\bar \gamma =20$dB),
the user can get a maximum gain of around $50\%$ of the 
achievable rate when the channel condition is
at its peak. On the other hand, at low SNR ($\bar \gamma =-10$dB), the achievable rate at peak
channel conditions can exceed four times its average value.
The intuitive suggestion of sorting the users in ascending order of their expected SNR 
is also supported by the numerical examples shown in Section~\ref{sec:numerical_results}.

In order to solve \eqref{eq:PF_main} numerically at the mobile terminal, 
$f_R(r)$ should be estimated from the continuously measured channel conditions 
(after every pilot signal transmitted by the base station). 
The practical implementation steps of channel statistics (PDF) estimators is out
of the scope of this work and was discussed in the signal processing literature.
As an example, the PDF estimation using order statistic filter bank
was suggested in \cite{SuEsRaVa:1994}.
 
A major concern in distributed systems in general is the effect of ill-behaving
mobile terminals. In our proposed proportional fair MUSwiD scheme,
the users obtain their thresholds locally. However, if one mobile terminal
uses lower threshold than its correct threshold, the performance of all
next users in the sequence will be affected and degraded.
We demonstrate here that it is possible to assign a monitoring task to the
base station in order to detect ill-behaving users without the need of knowing
the channel PDF of every user.
The suggested centralized monitoring mechanism works as follows;
The users compute their channel thresholds locally and
update the main scheduler at the base station about their thresholds. This does not produce
significant feedback load as the threshold values are re-computed only after
sound variations in the channel PDF.
The base station makes sure that the requested rates by the 
scheduled users are above their thresholds,
and it tracks two quantities (measures) for each mobile user
that are updated in real-time whenever the user has the opportunity to
request transmission (i.e.~condition \eqref{eq:no_flag_condition}):
(i) $R^c_i$: the average requested rate of user $i$ when 
condition \eqref{eq:no_flag_condition} is valid,
(ii) $P_i$: the success ratio of exceeding the channel threshold when 
condition \eqref{eq:no_flag_condition} is valid.
The base station makes sure that the measured quantities are
consistent with \eqref{eq:PF_main}.

In a mathematical context, $R^c_i$ is defined in \eqref{eq:conditional_rate} and $P_i$ is defined as
$P_i=\Pr\{r_i \geq r^*_i|\mathbf{r} \in \mathcal{S}_i\}$.
Note that tracking $R^c_i$ and $P_i$ does not 
require any additional feedback load. The base station can
detect a wrongly used threshold if the following condition is true:
\begin{equation}
\label{eq:detecting_function}
\left|\frac{r^{*}_{i} \; (1-P_i)}{R^c_i}-(M-i)\right|>\epsilon,
\end{equation}
where $\epsilon$ is the tolerance value for the accuracy of achieving
condition \eqref{eq:PF_main}.

\section{performance Analysis -- Comparison with Full-Feedback Schemes}
\label{sec:performance_analysis}
We provide different numerical examples to compare the performance of MUSwiD scheduling schemes
with the performance of full-feedback MUSelD scheduling schemes. We briefly summarize the achievable rates
using MUSelD schemes which were studied in the literature such as 
in \cite{YaAl:2006} and \cite{ShGoTh_eusipco:2009}.

\subsection{Review of Achievable Rates of Full-Feedback Selection-Based Systems}
\label{sec:recap_selection}

In MUSelD scheduling schemes, the users continuously update the centralized
scheduler at the base station about their instantaneous achievable rates $r_i$,
and the scheduler chooses the user that maximizes the scheduler metric.
There are many scheduling metrics suggested in the literature.
A survey and comparison between different schemes is provided in \cite{ShGo3:2008}.
In a generic form that enables achieving all Pareto-optimal points, the scheduling
criterion is to select a user $m$ with maximum weighted rate metric \cite{WaGiMa:2007},
i.e.~
$m=\arg \max_i \mu_i r_i$,
where $\mu_i$ is a weighting factor assigned to user $i$. 
In full feedback MUSelD scheduling, the expected achievable rates 
in terms of ($f_{\Gamma}(\gamma)$) 
is known from the literature (e.g.~\cite{ShGo3:2008}, \cite{ShGoTh_eusipco:2009}):
\begin{equation}
\label{eq:rate_MUSelD_SNR}
R_i= \int_{0}^\infty f_{\Gamma_i}(\gamma)  \; . \; \prod_{j\not =i}F_{\Gamma_j}\left((1+\gamma)^{\frac{\mu_i}{\mu_j}}-1\right) 
\; . \; \log(1+\gamma) d\gamma .
\end{equation}

The average channel access ratio (percentage of being scheduled) is:
\begin{equation}
\label{eq:ar_MUSelD_SNR}
\text{AR}_i= \int_{0}^\infty f_{\Gamma_i}(\gamma)  \; . \; 
\prod_{j\not =i}F_{\Gamma_j}\left((1+\gamma)^{\frac{\mu_i}{\mu_j}}-1\right) \; d\gamma .
\end{equation}

We present \eqref{eq:rate_MUSelD_SNR} and \eqref{eq:ar_MUSelD_SNR} in an alternative form 
using the PDF of the achievable rates $f_{R}(r)$:
\begin{equation}
\label{eq:rate_MUSelD}
R_i= \int_{0}^\infty r \; f_{R_i}(r)  \; \prod_{j\not =i}F_{R_j}\left(\frac{\mu_i \; r}{\mu_j}\right) \;  dr ,
\end{equation}

\begin{equation}
\label{eq:ar_MUSelD}
\text{AR}_i= \int_{0}^\infty f_{R_i}(r)  \; \prod_{j\not =i}F_{R_j}\left(\frac{\mu_i \; r}{\mu_j}\right) \;  dr .
\end{equation}

\subsection{Network Models and Fairness Measures}
\label{sec:network_models}
We compare MUSwiD and MUSelD schemes using different network scenarios (in terms
of the distribution of the expected channel conditions of the users)
and for different number of users.
We analyze the case of independent and identically distributed (i.i.d.) Rayleigh
block-faded channels as well the case of independent and non-identically distributed
Rayleigh channels which is more realistic from practical perspective.
We provide numerical examples for the asymmetric channel distribution case using two models: 
\begin{subequations}\label{eq:network_models}
\begin{align}
& \text{Model 1: }  \bar \gamma_i=\gamma_{\text{min}}+(2i-1).\frac{\gamma_{\text{max}}-\gamma_{\text{min}}}{2 M} \label{eq:network_model_1}\\
& \text{Model 2: }  \bar \gamma_i=\left[\sqrt{\gamma_{\text{min}}}+
\frac{2i-1}{2M}.\left(\sqrt{\gamma_{\text{max}}}-\sqrt{\gamma_{\text{min}}}\right)\right]^2, \label{eq:network_model_2}
\end{align}
\end{subequations} 
where $\gamma_{\text{max}}$ and $\gamma_{\text{min}}$ in \eqref{eq:network_model_1} 
and \eqref{eq:network_model_2} define respectively the upper and lower limits for the average SNR in the network.
We used in our numerical results 20 dB and 0 dB respectively.

We compare two variants of the scheduling criteria: (i) the maximum sum achievable rate,
and (ii) our proposed proportional fair scheduler.
We use two performance measures in our comparisons: (i) the sum achievable rate
in the network, and (ii) the degree of fairness (DOF) among the users.
There are several fairness measures suggested in the literature.
We opt in this work to use the well-known Jain's fairness index \cite{JaChHa:1984}: 
\begin{equation}
\label{eq:Jain_index}
\text{DOF}\equiv\frac{\left(\sum_{i=1}^Mx_i\right)^2}{M\sum_{i=1}^M x_i^2},
\end{equation}
where $x_i$ is a user-related metric. In our numerical examples we used two 
metrics for $x_i$:
\begin{itemize}
\item Resource sharing fairness: $x_i$ is selected to be the expected channel 
access ratio $\text{AR}_i$. 
\item Multiuser diversity gains fairness: we propose the following metric 
as well for the fairness measure:
\begin{equation}
\label{eq:my_fairness_metric}
x_i\equiv\frac{R_i}{\int_{0}^\infty r \; f_{R_i}(r) dr},
\end{equation}
where $R_i$ is the achievable rate of user $i$ according to the applied
scheduling scheme.
\end{itemize}

\subsection{Numerical Results}
\label{sec:numerical_results}

\begin{figure}[htb]
 \begin{center}
 \includegraphics[scale=0.6]{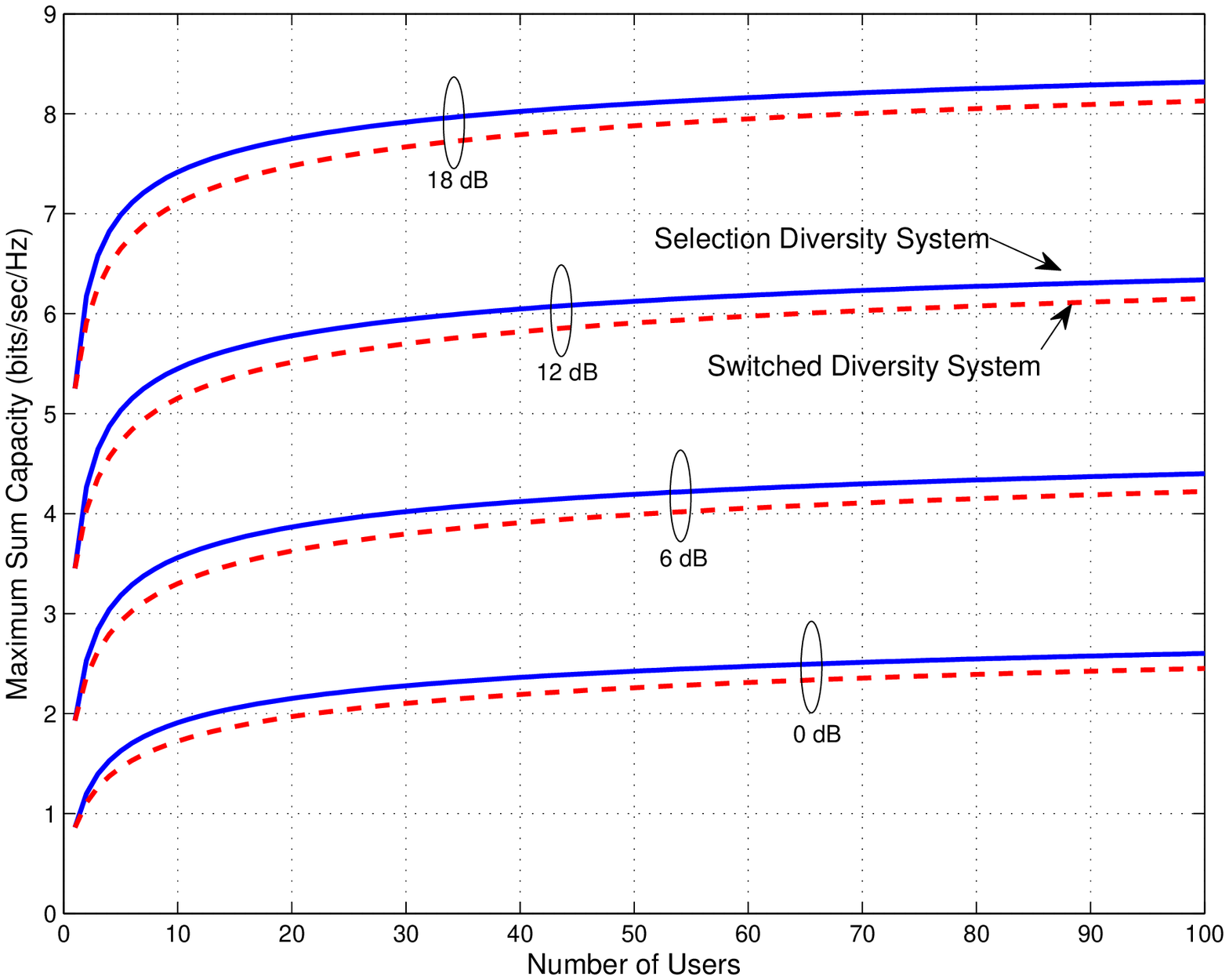}
\caption{Maximum sum achievable rate (capacity) comparison between the selection diversity system
(solid blue lines) and the switched diversity system (dashed red lines) as a function
of the number of users over i.i.d. Rayleigh block-fading channels. Results are based on average
SNR of 0, 6, 12 and 18 dB.}
   \label{fig:Sum_Cap_iid}
 \end{center}
\end{figure}

\begin{figure}[htb]
 \begin{center}
 \includegraphics[scale=0.7]{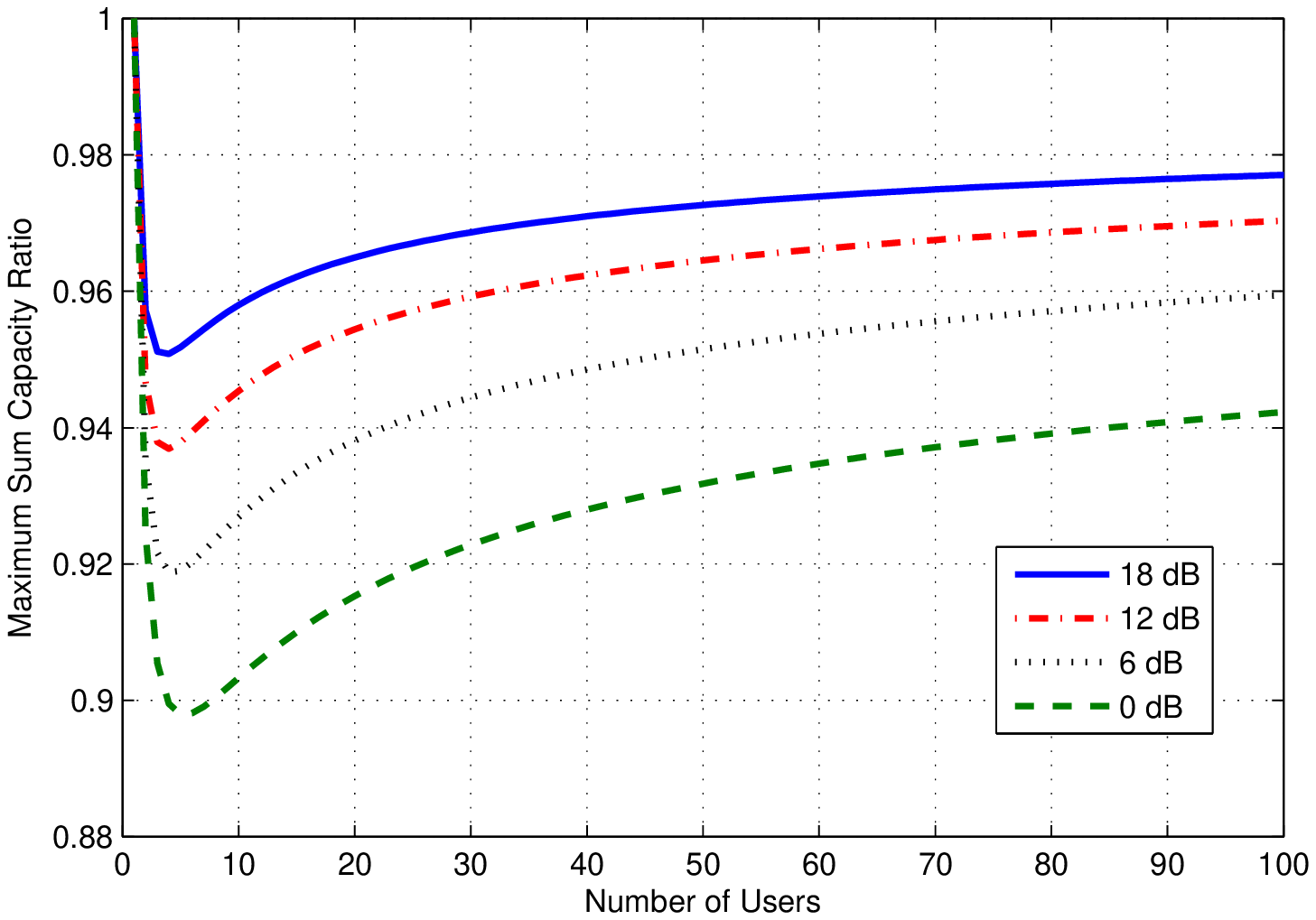}
\caption{The ratio between the maximum sum achievable rate (capacity) of switched diversity system
and the maximum sum capacity of the selection diversity system as a function of the number
of users over i.i.d. Rayleigh block-fading channels.}
   \label{fig:Sum_Cap_Ratio_iid}
 \end{center}
\end{figure}

\begin{figure}[htb]
 \begin{center}
\includegraphics[scale=0.65]{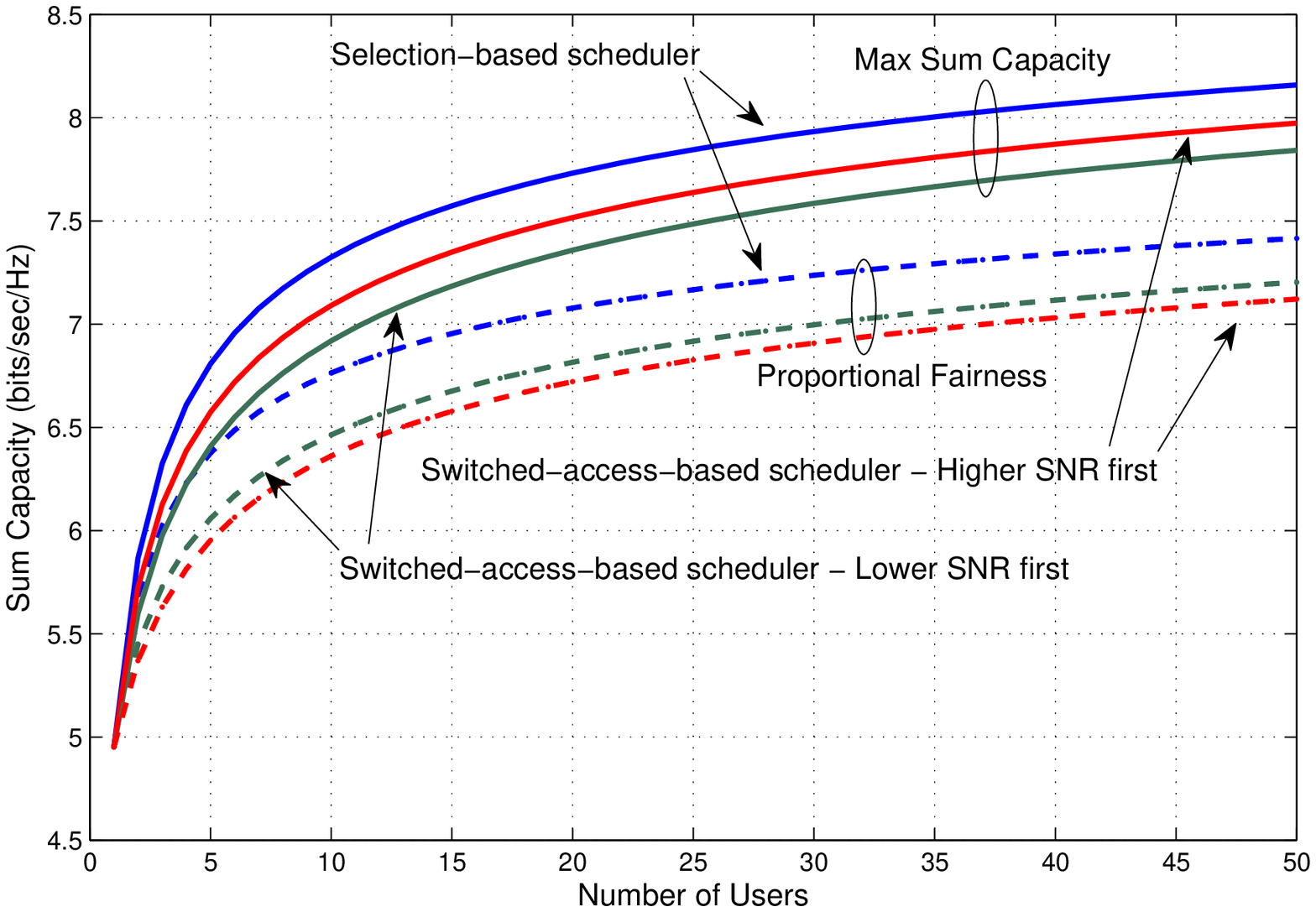}
\caption{Sum achievable rate comparison between the selection diversity system
(blue lines) and the switched diversity system (red and green lines) as a function
of the number of users for maximum sum rate scheduling (solid lines) and proportional fairness
scheduling (dashed lines). The users have Rayleigh block-fading channels with average
SNR distributed according to \eqref{eq:network_model_1}. Two feedback sequence strategies
are examined: ascending (green lines) and descending (red lines) average SNR order.}
   \label{fig:Sum_Cap_non_iid}
 \end{center}
\end{figure}

\begin{figure}[htb]
 \begin{center}
\includegraphics[scale=0.65]{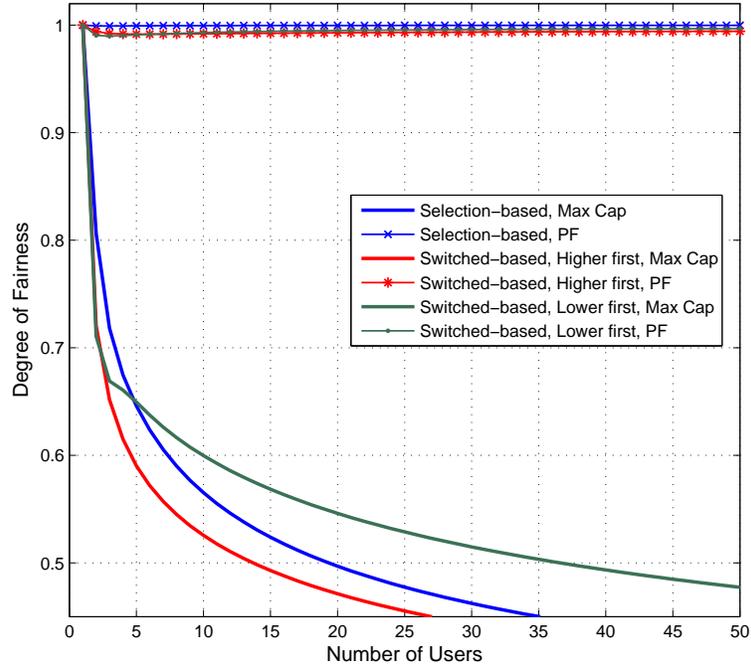}
\caption{Fairness measure by applying Jain's index \eqref{eq:Jain_index} with
($x_i=\text{AR}_i$) the average channel access ratio. The users have 
Rayleigh block-fading channels with average
SNR distributed according to \eqref{eq:network_model_1}.}
   \label{fig:ar_fairness}
 \end{center}
\end{figure}

\begin{figure}[htb]
 \begin{center}
\includegraphics[scale=0.65]{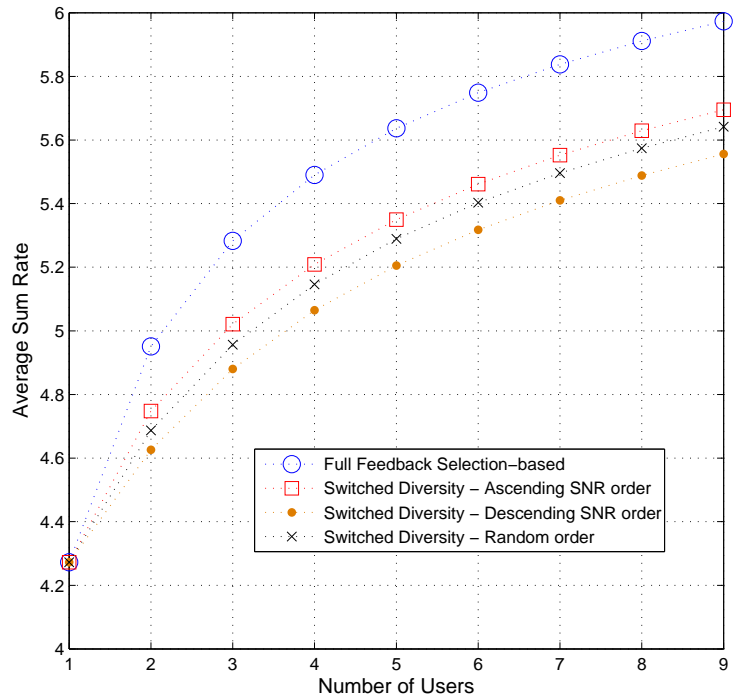}
\caption{Sum achievable rate comparison between the selection diversity system
and the switched diversity system as a function
of the number of users for proportional fairness scheduling. 
The users have Rayleigh block-fading channels with average
SNR distributed according to \eqref{eq:network_model_2}.}
   \label{fig:Sum_Cap_non_iid_2}
 \end{center}
\end{figure}

\begin{figure}[htb]
 \begin{center}
 \includegraphics[scale=0.65]{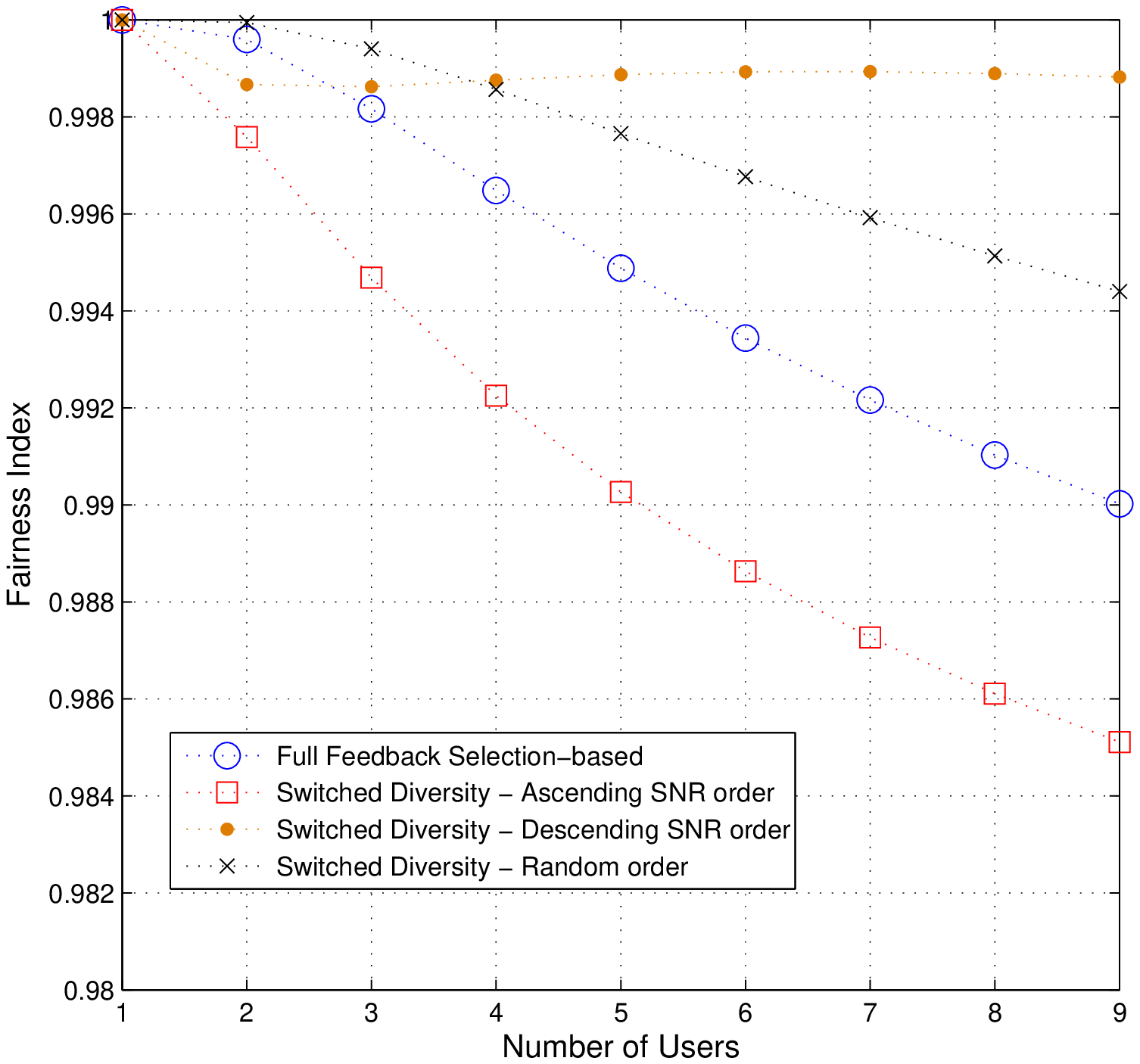}
\caption{Fairness measure by applying Jain's index \eqref{eq:Jain_index} with
\eqref{eq:my_fairness_metric}. The users have Rayleigh block-fading channels with average
SNR distributed according to \eqref{eq:network_model_2}.}
   \label{fig:my_fairness}
 \end{center}
\end{figure}

Fig.~\ref{fig:Sum_Cap_iid} shows the comparison between MUSwiD and MUSelD schemes
under i.i.d. Rayleigh block-fading conditions for different values of the identical average SNR
of the users. The feedback sequence of the MUSwiD scheme is irrelevant in this case as the
channels are identical. The maximum sum achievable rates are used for the comparison.
The sum capacity\footnote{In the special case of i.i.d. Rayleigh block-fading channels with identical average SNR,
the maximum sum capacity of the MUSelD can be computed as \cite{ShGoTh_eusipco:2009}:
\begin{displaymath}
\sum_{i=1}^M R_i=\sum_{i=1}^{M}(-1)^{(i-1)}
 \binom{M}{i}
\exp\left(\frac{i}{\bar{\gamma}}\right)
E_1\left(\frac{i}{\bar{\gamma}}\right),
\end{displaymath}
where $E_1$  is the exponential integral function \eqref{eq:E1_definition}.}
were computed using \eqref{eq:sum_capacity_iid_MUSwiD} for the MUSwiD scheme where the
per-user thresholds optimization follows \eqref{eq:thresholds_mu_SNR}.
Fig.~\ref{fig:Sum_Cap_iid} shows that switched-diversity scheduling schemes 
operate within 0.3 bits/sec/Hz from the ultimate network capacity of 
full feedback systems in Rayleigh block-fading conditions over wide range of SNR and for any number
of users. This rate loss is compensated by significant savings in the CSI feedback load.
At high SNR conditions, the ratio between
the sum capacity (i.e.~achievable rates) of MUSwiD schemes with respect to the sum capacity of MUSelD schemes decreases
as shown in Fig.~\ref{fig:Sum_Cap_Ratio_iid}. 
Fig~\ref{fig:Sum_Cap_non_iid} and Fig.~\ref{fig:ar_fairness} show
a comparison in terms of sum capacity and degree of fairness under
asymmetric channel conditions according to \eqref{eq:network_model_1}.
Both maximum sum capacity and proportional fairness are used in the comparison.
We used the derived analytical formulas in this paper to calculate 
the achievable rates for MUSwiD schemes.
Another numerical example is provided in  
Fig.~\ref{fig:Sum_Cap_non_iid_2} and Fig.~\ref{fig:my_fairness} 
for the proportional fair scheduler under the assumption of 
asymmetric channel distribution according to \eqref{eq:network_model_2}.
The fairness results in Fig.~\ref{fig:ar_fairness} are based on using
$x_i=\text{AR}_i$ in \eqref{eq:Jain_index}, while 
\eqref{eq:my_fairness_metric} is used for the fairness measure in Fig.~\ref{fig:my_fairness}.

The results in this section support the key messages learnt by studying the achievable
rate region in Section~\ref{sec:motivation} and demonstrate that they are valid for higher 
number of users. The performance of switched diversity is always within
0.3 bits/sec/Hz from the ultimate performance of full feedback schedulers. 
This is true for both maximum sum capacity and proportional fairness. 
The proportional fair scheduler provides very high degree of fairness
regardless of the used feedback sequence. The differences in fairness measures
of different feedback sequences  are negligible. 
The assessment of the performance of the sequence strategies is better judged
based on achievable rates which demonstrates that
sorting the users in a descending average SNR order is better for maximizing the sum achievable rate, 
while the opposite sequence is better for the 
proportional fair scheduler. These results are consistent with our results
in section~\ref{sec:motivation} and with our perceptive analysis in section~\ref{sec:proportional_fairness}.

\section{Conclusions}
\label{sec:Conclusions}
In this paper, we have proposed novel reduced-feedback scheduling schemes
that provides significant reduction of channel state information
feedback load at the cost of slight reduction in the
achievable multiuser diversity gains.
Our proposed schemes are based on the concept of multiuser switched diversity
that has been recently introduced in the literature.
We have provided rigorous mathematical treatment to analyze the 
performance of switched diversity scheduling schemes as well as
to optimize their performance.
We have also characterized the achievable rate region of these
scheduling schemes and provided a case study to understand their
main attributes and useful design options.
We proposed a proportional fair scheduler that overcomes major
technical challenges of the state-of-the-art proposals in the field.
Mainly, our proposed scheduler maintains fairness among users
and interestingly enables simpler optimization procedure.
we have demonstrated that, unlike other schedulers, the optimization procedure of
our proposed proportional fair scheduler can be distributed among the users.
We have shown that the distributed optimization mechanism can be supported
by a monitoring mechanism of the base station that enables the detection of
ill-behaving users based on real-time performance measurements.
Due to their features and performance, multiuser switched diversity scheduling systems
are actually attractive options for practical implementation in emerging
mobile broadband communication systems.


\bibliographystyle{IEEEbib}
\bibliography{./BIBFILES/papers,./BIBFILES/books}

\end{document}